\documentclass[prd,a4paper,showkeys,nofootinbib]{revtex4}
\usepackage{graphicx}

\begin{document}

\newcommand{\be}{\begin{equation}}
\newcommand{\ee}{\end{equation}}
\newcommand{\bq}{\begin{eqnarray}}
\newcommand{\eq}{\end{eqnarray}}
\newcommand{\bc}{\begin{center}}
\newcommand{\ec}{\end{center}}
\newcommand{\pd}{\partial}
\newcommand{\Mpc}{{\mathrm{Mpc}}}

\title{Gravitational shocks as a key ingredient of Gamma-Ray Bursts}

\author{Anastasios Avgoustidis\footnote{E-mail: a.avgoustidis@damtp.cam.ac.uk}$^{a}$, Raul Jimenez\footnote{E-mail: raul.jimenez@icc.ub.edu}$^{b,c}$, Luis \'Alvarez-Gaum\'e\footnote{E-mail: Luis.Alvarez-Gaume@cern.ch}$^d$, Miguel A. V\'azquez-Mozo\footnote{E-mail: Miguel.Vazquez-Mozo@cern.ch}$^e$}
\affiliation{$^a$School of Physics and Astronomy, University of Nottingham, University Park, 
Nottingham NG7 2RD, UK\\
$^b$Institucio Catalana de Recerca i Estudis Avancats (ICREA), 23 Passeig Lluis Companys, Barcelona, Spain \\
$^c$Institut de Ciences del Cosmos (ICC), Universitat de Barcelona (IEEC-UB), Marti i Franques 1, Barcelona 08028, Spain \\
$^d$Theory Group, Physics Department, CERN,\\ CH-1211 Geneva 23, Switzerland\\
$^e$Departamento de F\'isica Fundamental, Universidad de Salamanca,\\ Plaza de la
Merced s/n, E-37008 Salamanca, Spain}

\date{\today}

\begin{abstract}
We identify a novel physical mechanism that may be responsible for energy release in $\gamma$-ray bursts.  Radial perturbations in the neutron core, induced by its collision with collapsing outer layers during the early stages of supernova explosions, can trigger a gravitational shock, which can readily eject a small but significant fraction of the collapsing material at ultra-relativistic speeds.  The development of such shocks is a strong-field effect arising in near-critical collapse in General Relativity and has been observed in numerical simulations in various contexts, including in particular radially perturbed neutron star collapse, albeit for a tiny range of initial conditions.  Therefore, this effect can be easily missed in numerical simulations if the relevant parameter space is not exhaustively investigated.  In the proposed picture, the observed rarity of $\gamma$-ray bursts would be explained if the relevant conditions for this mechanism appear in only about one in every $10^4-10^5$ core collapse supernovae. We also mention the possibility that near-critical collapse could play a role in powering the central engines 
of Active Galactic Nuclei. 
\end{abstract}


\maketitle

\section{Introduction}

Gamma-ray bursts (GRBs) are short duration events, releasing up to $10^{51}-10^{53}$ ergs of electromagnetic radiation at 100 keV to 1 MeV, within a few seconds.  Pumping as much electromagnetic energy as the total output of the Sun during its whole lifetime, these events completely outshine all other $\gamma$-ray sources in the sky during their short durations.  They are however rare, occurring at an approximate rate of one burst per million years per galaxy\footnote{This estimate assumes bursts are isotropic.  If beamed the rate is larger by a factor of $2\pi/\theta^2$, with $\theta$ the beaming angle. It also assumes the rate does not depend on cosmological evolution.}.  The observed spatial distribution of GRBs has a high degree of isotropy indicating a cosmological origin~\cite{FenEpsetal}, and indeed observations of their associated afterglows yields cosmological redshifts, in some cases exceeding the ``dark ages barrier'' $z\simeq 6$.  Burst durations $T$ range from $10^{-3}$ to $10^{3}$ seconds and are known to roughly follow a bimodal distribution with ``short'' bursts of $T \lesssim 2\ {\rm seconds}$ and ``long'' ones of $T \gtrsim 2\ {\rm seconds}$~\cite{Kouvetal}, often with significant substructure.  Indeed, light-curves vary drastically from one burst to another, ranging from smooth to highly variable, with variability timescales down to milliseconds.  

The variability timescales observed provide a light travel time argument that leads to an estimated minimum size for their central engines of order tens of kilometers, typical of stellar mass neutron stars and black hole event horizons.  Thus, although the exact nature of their central engines remains largely unknown, they are believed to be linked with cataclysmic stellar events, with long burst conjectured to be caused by rare core collapse explosions of massive stellar progenitors (collapsar, hypernova~\cite{Woosley93,WooslMacFad,Paczynski}), and short ones believed to be linked to compact mergers of neutron star-neutron star (NS-NS mergers) or black hole-neutron star (BH-NS mergers)~\cite{LR-R}.  NS-NS merger type scenarios are based on phenomena that are independently observed and are believed to occur at approximately the correct rate, having short durations and containing enough binding energy to feed short GRBs.  On the other hand, collapsar scenarios receive observational support by the fact that long bursts often occur in star formation regions and have been associated with observed supernova (SN) explosions on several occasions (e.g.~\cite{Stanek,Hjorth,Galama98}).  Note, however, that this ``standard" GRB picture has recently been challenged by new strong bounds on the emitted neutrino flux set by the IceCube 
collaboration~\cite{IceCube_no_nus}.

The release of such large quantities of $\gamma$-radiation\footnote{This is implicitly assuming that the observed $\gamma$-ray photons directly originate from the engine.  One could imagine scenarios in which the energy is released in the form of some other weakly interacting particle, which gets converted to $\gamma$-rays during its flight towards the observer (see for example~\cite{Loeb}).} within the small volumes implied by the observed GRB variabilities and in the short time scales implied by GRB durations, generically gives rise, through the production of $e^+e^-$ pairs, to an opaque photon-lepton fireball~\cite{Goodman86,Paczynski86}, akin to the early universe.  However, such a fireball would naively produce a (quasi-)thermal spectrum, while the observed spectrum is clearly non-thermal indicating an optically thin source.  In particular, the observed low-energy spectrum is consistent with synchrotron emission from relativistic electrons~\cite{Katz94}, providing direct evidence for relativistic shocks in GRB's.  Since a relativistic particle flow is the generic outcome of a fireball, the general strategy behind fireball models is to convert this energy into a relativistic flow and then, at a later stage, reconvert it back into radiation through the development of shocks as the ejecta collide with the circumburst medium (external shock models~\cite{external}) or, perhaps more likely, as different shells of ejecta collide with each other due to the internal source variability (internal shock models~\cite{internal}).     

Though a complete model for long bursts is still lacking, a consistent picture is starting to emerge in terms of such a four-stage ``fireball shock scenario'' involving an inner engine that produces a (possibly highly variable) relativistic flow, an energy transport stage, a subsequent stage in which this energy gets converted into the observed radiation, and a final stage converting the remaining energy into a softer wavelength, long afterglow\footnote{For reviews see for example~\cite{Piran04,Meszaros06}.}.  However, even leaving aside possible tension with observations (e.g.~\cite{IceCube_no_nus}), major problems on the theoretical side remain, namely in identifying the nature of the central engine itself, and, in particular, understanding how the available energy can be efficiently channelled into a relativistic flow.  Indeed, a fundamental problem in the fireball picture is to understand the physical processes that lead to fireball formation in the first place: most popular models based on neutrino-antineutrino annihilation seem to fail by several orders of magnitude~\cite{JankRuf}.  This has led, for example, to the development of alternative models with super-strong magnetic fields and fast rotation~\cite{Vietri,MeszRees}.  It is clear that at least for some of the observed GRBs, a novel explosive process is required.  SN explosions provide of order $10^{51}$ ergs over timescales of at least weeks, while for large redshift GRBs up to $10^{54}$ ergs (over much shorter timescales) are needed. Thus, even for jetted GRBs with a lower energy requirement in the region of a few$\,\times\,10^{52}$ ergs, ordinary SN can marginally provide the energy needed. Most importantly, even though a number of mechanisms have been proposed to accelerate mildly relativistic flows (for example through propagation in a rapidly declining atmosphere), there is at present no well-understood mechanism able to produce an ultra-relativistic flow of a significant fraction of a solar mass of baryons to the required Lorentz factors.

In this note we point out a novel gravitational -- rather than hydrodynamic -- mechanism that may provide the missing key ingredient for understanding the physics of GRBs.  In this picture, the collision of collapsing stellar layers with a neutron core can induce a gravitational shock, which can in turn eject a small but appreciable amount of the collapsing material at ultra-relativistic Lorentz factors.  The proposed mechanism is based on type II critical black hole collapse, a strong-field effect in General Relativity that has been observed numerically on several, mostly spherically symmetric\footnote{For a discussion of cases beyond spherical symmetry, including the introduction of angular momentum and the axisymmetric problem, see \cite{reviews}.}, collapse simulations~\cite{Choptuik,NeilsChopt,Novak,NoblChopt}.  In the context of the initial value problem in General Relativity, such (exactly) critical solutions are known to occur for infinitely fine-tuned initial data that in fact form a set of measure zero, and so they are not expected to be physically realised in nature\footnote{They also suffer from naked singularities.}.  However, the effects considered in this paper arise for near-critical data corresponding to a small but finite fraction of initial conditions, making this mechanism relevant for explaining rare events like GRBs and, possibly, AGNs.

\section{Critical Type II Collapse in General Relativity\label{sec_collapse}}

Gravitational collapse is known to provide the engine behind many astrophysical phenomena. This is one of the reasons why black hole formation has remained a central field of study in General Relativity since the pioneering work by  Oppenheimer and Snyder \cite{oppy_snyder}. A specially interesting question is the existence of a threshold for the formation of a black hole as a function of the initial conditions for the gravitational collapse. Christodoulou's theorem 
\cite{christodoulou} for spherical collapse of a scalar field establishes that for sufficiently weak initial data the dynamical evolution leads to the dispersion of the collapsing matter without black hole formation, whereas for sufficiently strong initial data the system evolves into the formation of a black hole.  Following Christodoulou's work, Choptuik \cite{Choptuik} started a program to study numerically the gravitational collapse of different types of matter in various setups. To avoid the difficulties derived from the emission of gravitational waves in the collapse process, these studies were mostly restricted to spherical symmetry, although axially-symmetric situations were also analyzed.

The most interesting outcome of these studies was the discovery of critical gravitational collapse \cite{Choptuik} (see \cite{reviews} for a comprehensive overview of the subject). One considers a family of initial conditions for the collapse problem parametrized by a real number $p$, such that for $p\ll 1$ the time evolution leads to the complete dispersion of the infalling matter, whereas for $p\gg 1$ the collapse results in black hole formation. Numerical simulation showed the existence of a threshold value of the parameter $p=p^{*}$ such that a black hole is formed only when $p>p^{*}$. Moreover, in \cite{Choptuik} it was found that for $p\gtrsim p^{*}$ the radius of the forming black hole obeys a scaling relation
\begin{eqnarray}
r_{\rm BH}\sim (p-p^{*})^{\gamma}.
\label{scaling_rel}
\end{eqnarray}
What is remarkable about this scaling is that the critical exponent $\gamma$ is independent of the particular family of initial conditions chosen for the simulation. Further studies showed that its value only depends on the type of collapsing matter, as well as on the symmetry of the problem. 

The universality of the critical exponent in the scaling relation (\ref{scaling_rel}), together with the fact that the black hole at threshold has ``zero mass'', strongly suggests an analogy with type II critical phenomena. This justifies the name type II given to this kind of critical gravitational collapse. In fact, the analogy goes even further, since close to the forming singularity the geometry is approximately self-similar (either discrete or continuous). This allows a phase-space picture where the codimension-one surface $p=p^{*}$ can be seen as a critical surface separating the set of initial conditions leading to the formation of a black hole from those that evolve to Minkowski space-time after the dispersion of the infalling matter to infinity. Initial conditions on the critical surface $p=p^{*}$ evolve into a critical solution with either discrete or continuous self-similarity. This fixed point corresponds to an (unphysical) critical solution that describes a naked singularity which is unstable under perturbations orthogonal to the critical surface.

These properties contrast with the ones of the so-called type I critical gravitational collapse, in which the black hole that forms at $p=p^{*}$ has a nonzero size which is universal, again in the sense of being independent of the particular family of initial conditions of the collapse. One interesting question is what determines that a given system undergoes type I or type II critical behavior. On general grounds one can argue that type I takes places whenever there is a relevant mass scale in the problem and this scale dominates the dynamical evolution. This means that type II gravitational collapse can take place even in the presence of a mass scale, provided this scale is dynamically irrelevant, as it happens when the evolution of the system is dominated by the kinetic energy. This point will be crucial for our discussion in the following sections.

Although the actual simulation of the system is technically very involved, the physical picture of matter undergoing near-critical type II gravitational collapse is simple to describe. A configuration of matter starts collapsing under the influence of its own gravity with a certain profile of initial velocities. If these initial conditions are near-supercritical (i.e. $p\gtrsim p^{*}$) the evolution will produce a black hole containing a significant fraction of the total ADM mass of the space, while the rest of the matter is ejected to infinity at relativistic speeds. If, on the other hand, the system is slightly subcritical ($p\lesssim p^{*}$) the curvature at the center of the collapse grows to a maximum value before the matter is expelled to infinity, leaving flat space behind. Although no black hole is formed in this case, the maximum curvature at the center obeys the following scaling relation \cite{garfinkel_duncan}
\begin{eqnarray}
R_{\rm max}\sim (p^{*}-p)^{-2\gamma},
\end{eqnarray}
where $\gamma$ is the same critical exponent that characterizes the scaling of the size of the black hole for supercritical initial conditions. 

Critical gravitational collapse has been a central subject in numerical General Relativity over the last two decades. Although most of the numerical analysis carried out so far has been restricted to highly symmetric situations, it is important to stress that this kind of physical phenomena is not of just purely academic interest. For example, there is strong evidence that critical behaviour is robust with respect to the presence of angular momentum in the collapse \cite{angmomen}. In spite of this, the application of critical gravitational collapse to astrophysical situations has remained largely unexplored, apart from a few exceptions \cite{Novak98,Novak,NoblChopt}. These works have been focused on the study of black hole formation resulting from the collapse of a neutron star that accretes an outer layer of matter. Numerical simulations showed that there is a window in the parameter space of the problem for which type II critical gravitational collapse takes place.

\section{Near-critical collapse as an engine for GRB\scriptsize{s}}

The striking results of~\cite{Novak,NoblChopt} may have non-trivial implications for cataclysmic core collapse explosions and in particular for long GRBs that are believed to be associated with them.  The collision of collapsing outer stellar layers with the central neutron core that can take place in SN explosions, provides a perturbing agent that can set up the initial conditions considered in those numerical studies.  The author of \cite{Novak} finds that for large neutron stars or central densities an inward radial velocity of the outer neutron star layers as low as $0.1c$ can produce critical behaviour with the ensuing development of a gravitational shock.  For stable neutron stars -- isolated or in a binary system -- it is difficult to envision how such an inward velocity could be injected, and, even in the case of unstable neutron star solutions in the context of scalar-tensor gravity, the velocity induced cannot be larger than a few percential units of $c$~\cite{Novak98,Novak}.  However, for cataclysmic stellar collapse, such an inward velocity can be induced on the neutron core through kinetic energy transfer from the nearly free-falling outer layers that are moving with velocities as large as $\sim 0.3c$.  

\begin{figure}[h]
  \begin{center}
    \includegraphics[height=3.5in,width=4in]{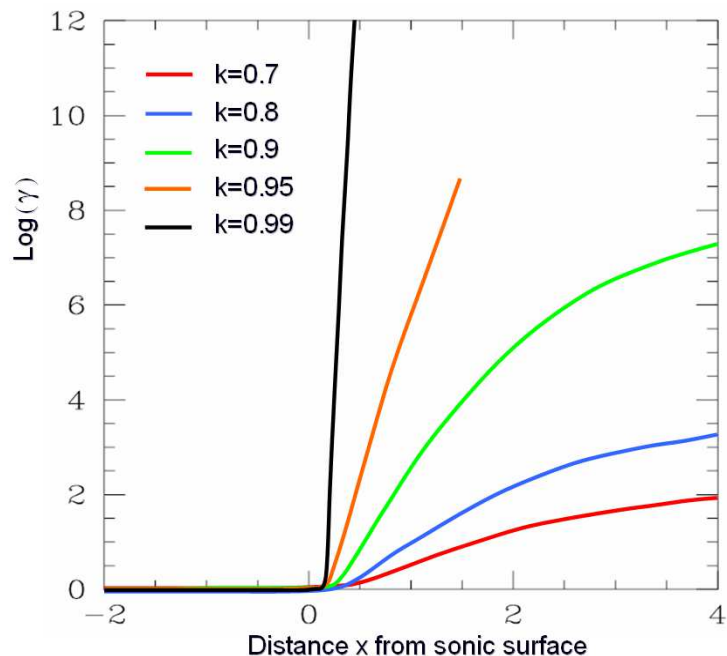}
    \caption{\label{gammas}  Ejecta Lorentz factors at criticality plotted against distance from the shock surface, for different choices of equations of state $k=p/\rho$.  The shock surface is at $x=0$.  Adapted from \cite{NeilsChopt}.}
   \end{center}
  \end{figure}

In the numerical studies of Refs.~\cite{Novak,NoblChopt} a gravitational shock can develop, which leads to the ejection of a fraction of the collapsing material at ultra-relativistic Lorentz factors.  The fraction of the ejected material depends on the initial conditions of the numerical experiment -- in particular reaching unity at criticality as explained above -- , while the maximal ejection velocity depends strongly on the equation of state (Fig.~\ref{gammas}).  Similar critical behaviour has been also found in other perfect fluid collapse simulations~\cite{EvCol,KoHaAd,HarMae,NeilsChopt}, where the ejecta velocity gets highly relativistic for stiff equations of state~\cite{NeilsChopt}, thus making this mechanism particularly interesting for neutron stars and, in this case, neutron cores.  

Although the exact critical solution (leading to the formation of a ``zero-mass black hole'' with a naked singularity) is unphysical and, having zero measure, does not occur in nature, near-critical data forming a finite-measure set in the manifold of initial conditions for the collapse problem can lead to the required behaviour.  In this case, rather than ejecting all the mass of the neutron star leaving behind a zero mass black hole, a certain fraction (depending on how far from criticality one is) of the available mass is ejected with a velocity distribution peaking at ultra-relativistic speeds, while the rest forms a black hole (super-critical case) or moves outwards too at a range of lower speeds carrying the bulk of the kinetic energy (sub-critical case).  The power-law scaling behaviour of equation (\ref{scaling_rel}) persists over a significant range of initial conditions around criticality for which a fraction as large as $30\%$ of the neutron star mass is ejected.  Ultra-relativistic velocities, however, occur for small deviations from criticality.  In the context of core collapse events considered here, the relevant range of initial conditions corresponds to the large velocities ($\gtrsim 0.1c$) which need to be induced on the outer layers of the neutron core from the collapsing stellar layers, and the closeness to spherical symmetry of the collapse.  The observed rare frequency of GRB events requires the fraction of the relevant near-critical parameter range over the prior to be of order $10^{-5}-10^{-4}$, but this fraction can 
be up to two orders of magnitude higher for jetted GRBs.          

To explore the plausibility of this mechanism and to estimate the required astrophysical parameters, we consider a simple model for the inelastic collision between two concentric spherical shells with masses $m_s, m_r$ and $\gamma_s, \gamma_r$ respectively.  Their collision leads to a single (merged) shell with Lorenz factor $\gamma_m$ and internal energy $\cal E$.  The relativistic conservation equations read:
\bq\label{cons_eqs}
\left\{ \begin{array}{l}
m_s \gamma_s + m_r \gamma_r = (m_s+m_r+{\cal E}/c^2)\gamma_m  \\
m_s \sqrt{\gamma_s^2-1} + m_r \sqrt{\gamma_r^2-1} = (m_s+m_r+{\cal E}/c^2) \sqrt{\gamma_m^2-1} \, .
\end{array} \right.
\eq       
In the case described above, where a rapidly collapsing shell $\left\{ m_r,\gamma_r\right\}$ hits the static outer shell of the core $\left\{m_s,\gamma_s=1\right\}$, equations (\ref{cons_eqs}) yield for the Lorentz factor of the merged shell:
\be\label{g_m}
\gamma_m=\frac{\gamma_r+\frac{m_s}{m_r}}{\sqrt{(\gamma_r+\frac{m_s}{m_r})^2-
(\gamma_r^2-1)}}\,.
\ee   
In particular, solving for the mass ratio $m_s/m_r$ yields one positive solution that can be expressed as:
\be\label{mass_ratio} 
\frac{m_s}{m_r}=\gamma_r \left(\frac{v_r}{v_m}-1 \right) \, ,
\ee   
where $v_r,v_m$ are the velocities corresponding to $\gamma_r,\gamma_m$ respectively.  Thus, for a collapsing stellar shell with $v_r\simeq 0.25c$ to induce a $v_m\simeq 0.1c$ on the static neutron core shell, the mass ratio must be $m_s/m_r\simeq 1.5$.  For a less massive neutron core shell, $v_m$ is larger with $v_m\rightarrow v_r$ as $m_s/m_r$ approaches zero (see equation (\ref{g_m})).  For equal shell masses, $m_s=m_r$, one has $\gamma_m=\sqrt{(\gamma_r+1)/2}$ corresponding to $v_m\simeq 0.13c$.  Since the neutron core is of order 1 $\rm M_\odot$, a stellar layer of mass $\gtrsim 1\ \rm M_\odot$ collapsing at the expected speeds of around $0.3c$ induces on the core a collapse velocity of $0.15c\lesssim v_m \lesssim 0.3c$.   

\begin{figure}[h]
  \begin{center}
    \includegraphics[height=3.7in,width=4in]{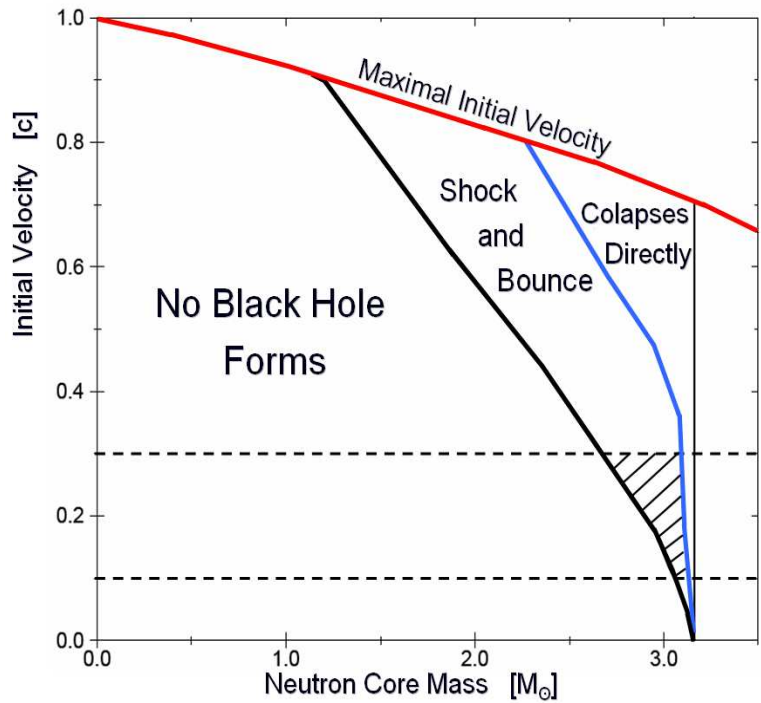}
    \caption{\label{par_space} Inward velocity amplitude versus mass of neutron core.  The part of the region of the parameter space leading to shock and bounce behaviour that can be relevant in our GRB picture is shaded.  Adapted from~\cite{Novak}.}
   \end{center}
  \end{figure}

This collision produces an internal energy of:
\be\label{internal}
{\cal E} = m_r c^2 \left[ \frac{\gamma_r+m_s/m_r}{\gamma_m}-\left( 1+\frac{m_s}{m_r} \right) \right] \, .
\ee   
For $m_s=m_r$ this yields ${\cal E}=m_r c^2 ( \sqrt{2(\gamma_r+1)}-2 )$, which for $v_r\simeq 0.25c$ gives ${\cal E}\simeq m_r v_r^2/4$, only about $1\%$ of the rapid shell's rest mass.  This energy could induce a small characteristic x-ray signal. 

The important point to highlight here is that the velocity induced on the core from this collision is of the same magnitude as that required by simulations to trigger a near-critical collapse event, which, as explained above, produces a gravitational shock at a finite distance from the core.  Thus, the material located outside the sonic surface is spherically expelled, with matter located near this surface developing ultra-relativistic Lorentz factors.  Sufficiently close to criticality, the required Lorentz factors of 300 or more can be generated, which is a novel feature of the (near-)critical collapse GRB scenario.  At the same time, layers of matter further away from the sonic surface will be ejected with a range of (lower) speeds.  The ultra-relativistically moving shell can then collide with slower, but still relativistic, shells as well as the interstellar medium (ISM), giving rise to a prompt and afterglow emission as in standard internal-external shock models~\cite{internal,external}.  Therefore, the required short-term variability can be produced through collisions of shells moving at different speeds and the total duration can reach up to tens (or hundreds) of seconds, much like in the standard picture.

Let us briefly review the basic elements of this picture.  The collision between the rapid and a slower shell are subject to the conservation equations (\ref{cons_eqs}), where the index `$s$' now stands for `slow'.  In this case $\gamma_r > \gamma_s \gg 1$ and the Lorentz factor of the merged shell is:       
\be\label{g_mf} 
\gamma_m\simeq \sqrt{2} \sqrt{\frac{m_r\gamma_r+m_s\gamma_s}{m_r/\gamma_r+m_s/\gamma_s}}=\sqrt{2\gamma_r\gamma_s}
\sqrt{\frac{\gamma_r/\gamma_s+m_s/m_r}{1+(m_s/m_r)(\gamma_r/\gamma_s)}}
\, .
\ee 
Clearly, for shells of comparable masses, $m_s\simeq m_r$, we have $\gamma_m\simeq \sqrt{2\gamma_r\gamma_s}$, so for $\gamma_r\simeq 300$, $\gamma_s\simeq 10$ the merged shell still moves ultra-relativistically with $\gamma_m\simeq 80$.  In this case the collision produces a huge amount of internal energy which, in the observer frame, reads:
\be\label{E_int}
U_{\rm int}\equiv \gamma_m {\cal E} = m_r c^2 \left[ \gamma_r + \gamma_s \frac{m_s}{m_r} - \left( 1+\frac{m_s}{m_r}\right)\gamma_m \right] \, .
\ee          
For the above values this gives about 150 times the rest mass of the rapid shell, so that about half of its original kinetic energy has been converted into internal (see for example~\cite{Piran04}).  The hydrodynamics is described by the Blandford-McKee solution~\cite{BlandMcKee,KobPirSar}.  There is a forward and a reverse shock with Lorentz factors depending on the corresponding Lorentz factors and the density ratio of the shells.  

Shock-heating is expected to produce, through Fermi acceleration, a power law distribution of random relativistic electrons, $N(\gamma_e)\propto \gamma_e^{-p}$ with $p\gtrsim 2$, above some minimum $\gamma_{e,{\rm min}}$.  This minimum random Lorentz factor of electrons is related to the bulk Lorentz factor $\Gamma$ between the shocked and unshocked fluid through $\gamma_{e,{\rm min}}\sim (m_p/m_e) \epsilon_e \Gamma$, where $m_p,m_e$ are the proton and electron masses respectively and $\epsilon_e\equiv U_e/U$ parametrises the fraction of internal energy which is in the form of `thermal' energy of electrons.  A large magnetic field (assumed to be randomly oriented) can be built up due to turbulent dynamo effects behind the shock~\cite{MeszRees93a,MeszRees93b}, so these electrons will synchrotron radiate.  Let $\epsilon_B\equiv U_B/U$ parametrise the ratio of magnetic field energy density to the internal energy.  Then, the magnetic field in the fluid frame is $B=\sqrt{8\pi\epsilon_B U}=c\Gamma\sqrt{32\pi \epsilon_B n m_p}$, where $n$ is the pre-shock particle density.  Here, the factor $\Gamma$ comes precisely from the relation between pre-shock and post-shock quantities, determined by the Blandford-McKee jump conditions.    

\begin{figure}[h]
  \begin{center}
    \includegraphics[height=3.7in,width=4in]{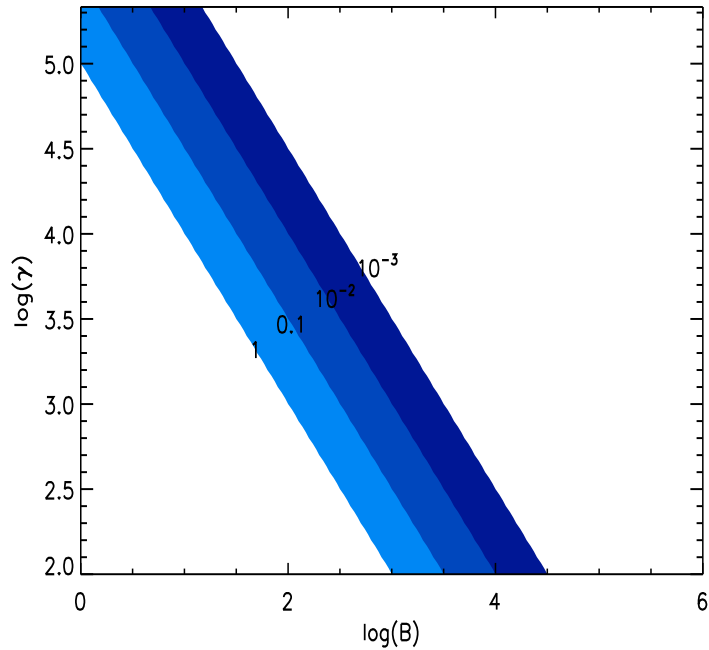}
    \caption{\label{masses} Mass of layer required to be ejected relativistically 
             by the shock as a function of magnetic field strength (in G) and random  
             Lorentz factor of the shocked electrons in the layer, for a GRB 
             emission power of $10^{52}\ {\rm erg\,s^{-1}}$.  The contours 
             correspond to 0.001, 0.01, 0.1 and 1 solar masses.}
   \end{center}
  \end{figure}

The peak synchrotron frequency in the observer frame is given by $\nu_{\rm peak}=(eB/2\pi m_ec)\gamma_e^2\Gamma$, where $e$ is the electron charge and $\gamma_e\sim \gamma_{e,{\rm min}}$.  The spectrum has a break at $\nu_{\rm peak}$ with $F_\nu\propto \nu^{1/3}$ for $\nu<\nu_{\rm peak}$ and $F_\nu\propto \nu^{-(p-1)/2}$ for $\nu>\nu_{\rm peak}$.  Matching this to observed spectra at high photon energies requires $p\simeq 2.5$.  For plausible parameter choices the emission can be in the required 100 keV range  with a duration (assumed to be determined by the hydrodynamic scale $t_{\rm hydr}\sim \Delta R/2c\Gamma^2$, where $\Delta R$ is the rapid shell's thickness) of order a few (tens of) seconds.  The corresponding observer frame synchrotron cooling time $t_{\rm cool}=\gamma_e m_e c^2/\Gamma P_{\rm sync}=3m_ec/4\sigma_TU_B\gamma_e\Gamma$, where $\sigma_T$ is the Thomson cross section, is in the (tens of) millisecond range, corresponding to a duty cycle of order $10^{-3}$~\cite{SarNarPir}.  
The synchrotron power per electron (averaging over electron orientations) is 
$P_e \simeq \frac{4}{9} \frac{e^4 B^2}{c^3 m_e^2} \gamma^2$, so for an observed GRB power of $P_{\rm GRB}$ one needs $N_e\simeq P_{\rm GRB}/P_e$ electrons of Lorentz factors of order $\gamma$.  Then, assuming an equal number of protons in the collapsing stellar layers, the shock needs to relativistically eject a thin layer of order $10^{-2}-10^{-1}\ {\rm M}_\odot$ or less in order to provide a power of $P_{\rm GRB}\simeq 10^{52}\ \rm{erg\, s^{-1}}$, for typical magnetic fields and Lorentz factors of random electrons (Fig.~\ref{masses}).

\section{Summary and Discussion}

It is important to note that the effects described in this paper do not correspond to generic or representative solutions of the collapse problem.  Rather, they are relevant for a small region of parameter space and can be easily missed in numerical simulations if this parameter space is not exhaustively sampled.  On the other hand, GRBs are observed to be equally rare so one should expect that the key physical mechanism behind them be not generic.   

There is direct evidence for relativistic shocks in GRBs, the low-energy spectrum being consistent with synchrotron emission from relativistic electrons. However, the physical mechanisms that could produce the required ultra-relativistic Lorentz factors, of order 100's or more, are not well-understood.  The gravitational shocks described here provide a novel mechanism, based on critical phenomena in non-linear gravity that are known to produce the required ultra-relativistic ejecta speeds.  These shocks are much more efficient in directly ejecting ultra-relativistic material (baryons) than their hydrodynamic counterparts.  Indeed, numerical work in the context of radially perturbed stable neutron stars~\cite{Novak,NoblChopt} demonstrates that, for stiff enough equations of state, the ejecta velocities at criticality are highly ultra-relativistic.  Solutions which are near-critical can then occur for a finite range of initial conditions, and the small fraction of initial data giving rise to violent stellar material ejection corresponds to the small observed frequency of GRB explosions. 

One is then left with the problem of identifying an astrophysical mechanism which can induce an inward velocity of order $0.1c$ or larger on a neutron star density object, thus setting up the relevant initial conditions for the studied collapse problem with its related near-critical phenomena.  For stable neutron stars (or even unstable ones in scalar-tensor gravity~\cite{Novak98}) this looks impossible, but in the context of cataclysmic core collapse events this may appear to proceed naturally.  Indeed, the collapsing iron core already develops significant velocities before it comes to a grinding halt when it reaches nuclear densities, while, in addition, significant amounts of kinetic energy can be transferred from the nearly free-falling outer layers to the newly formed neutron core.  These layers are known to develop inward velocities as large as $~0.3c$, which, as we have demonstrated, can easily induce a velocity greater than $0.1c$ on the outer neutron core.  This then provides the type of perturbations studied in~\cite{Novak,NoblChopt}, giving rise -- in a small region of parameter space -- to near-critical behaviour and the ensuing development of the gravitational shock described.     

As we have explained, this shock mechanism leads to the direct production of an ultra-relativistic flow with the required Lorentz factors of up to a few hundreds, as a thin layer of collapsing material near the sonic surface is ejected ultra-relativistically.  Nearby layers are also expelled, but at a range of lower (though still relativistic) speeds. This flow can then get converted into radiation via collisionless shocks as in the standard internal-external shock picture, thus reproducing the short-scale variability and total duration of long GRB events.  It is likely that the gravitational shock itself also liberates significant thermal energy that may contribute to the prompt emission.  Unfortunately, it is not possible to estimate the amount of this energy without a complete non-linear gravity simulation.   

Note that in this paper we have only discussed the spherical collapse case leading to isotropic ejecta, 
whereas there is strong evidence that (at least some) GRBs are jetted.  It is plausible that a jet structure 
could appear in the axisymmetric collapse problem, but unfortunately this case is not well-understood at present. 
Another possibility for supporting a jet structure is through a strong poloidal magnetic field in the spherical 
collapse model. It would be interesting to study this case in detail and investigate quantitatively the collapse 
problem in this context by performing high-resolution numerical simulations in full General Relativity.
              
Finally, note that the strong-field, non-linear nature of Type II (near-)critical collapse 
and the ensuing formation of a gravitational shock surface, render it an efficient 
mechanism for producing ultrarelativistic ejecta, which could be relevant also in 
other astrophysicsal phenomena. As reviewed in section \ref{sec_collapse} above, 
Type II collapse can arise when the rest mass energy of the collapsing matter is 
dynamically irrelevant, and, in particular, when the velocities involved are relativistic.  
It is therefore natural to look for situations in high-energy astrophysics where near-critical 
collapse could take place and play a role in generating ultra-relativistic flows.  An obvious
candidate is radio-jets in Active Galactic Nuclei (AGN). In the picture proposed in~\cite{crit_AGN},
 a mass of the order of $10^3\ \rm M_\odot$ of material collapsing relativistically could trigger 
a gravitational shock, ejecting a large percentage of the collapsing matter at relativistic speeds, 
and leaving behind a ``light" black hole.  In the presence of a poloidal magnetic field, the plasma 
collimates along two outgoing jets, and the associated electron synchrotron radiation could account 
for the observed radio luminosities, sizes and durations of  AGN jets.  For Lorentz factors of order 
100 and magnetic fields of a few hundred $\rm \mu G$, synchrotron electrons could shine for 
$10^6\ \mbox{yrs}$, producing jets of sizes of order $100\ \mbox{kpc}$.

\section*{Acknowledgements}
The authors would like to thank Scott Noble and Matt Choptuik for valuable correspondence, and Rosalba Perna for useful comments on this work.  The work of AA was supported by IEEC and FP7-PEOPLE-2007-4-3 IRG.  RJ was supported by FP7-PEOPLE-2007-4-3 IRG and by MICINN grant AYA2008-03531.  AA and RJ would like to thank CERN for warm hospitality while this work was being completed.

\end{document}